# Image Processing Based on Compound Flat Optics

You Zhou[1], Hanyu Zheng[2], Ivan I. Kravchenko[3], and Jason Valentine[4*]

[1]Interdisciplinary Materials Science Program, Vanderbilt University, Nashville, Tennessee 37212, USA
[2]Department of Electric Engineering and Computer Science, Vanderbilt University, Nashville, Tennessee 37212, USA
[3]Center for Nanophase Materials Sciences, Oak Ridge National Laboratory, Oak Ridge, Tennessee 37831, USA
[4]Department of Mechanical Engineering, Vanderbilt University, Nashville, Tennessee 37212, USA
[*]email: jason.g.valentine@vanderbilt.edu

## Abstract

**Image processing has become a critical technology in a variety of science and engineering disciplines. While most image processing is performed digitally, optical analog processing has the advantages of being low-power and high-speed though it requires a large volume. Here, we demonstrate optical analog imaging processing using a flat optic for direct image differentiation allowing one to significantly shrink the required optical system size. We first demonstrate how the image differentiator can be combined with traditional imaging systems such as a commercial optical microscope and camera sensor for edge detection. Second, we demonstrate how the entire analog processing system can be realized as a monolithic compound flat optic by integrating the differentiator with a metalens. The compound nanophotonic system manifests the advantage of thin form factor optics as well as the ability to implement complex transfer functions and could open new opportunities in applications such as biological imaging and machine vision.**

## Introduction

Imaging processing is a critical and rapidly advancing technology for various science and engineering disciplines with ever more complex digital tools opening the door for new avenues in biological imaging, 3D reconstruction, and autonomous cars. Edge-based enhancement[1,2] is particularly useful for data compression[3,4], object inspection[5], microscopy[6–8] and general machine vision[9]. Edge-based enhancement is accomplished using spatial differentiation that can be based on either electronic or optical architectures. While digital circuits are able to perform complex data processing there are limitations due to computation speed and power consumption. Optical analog computation[10], with the ability to process information directly using the optical signal, provides an alternative approach to perform large-scale and real-time data processing with minimal, if any, power consumption.

Traditionally, analog image differentiation has been performed using Fourier methods based on lenses and filter systems. However, the use of multiple conventional lenses, such as those found in 4*f* Fourier filtering, results in a large form factor which is not compatible with compact integrated systems. One option for significantly reducing the optical system size is to employ nanophotonic materials such as metasurfaces[11] and photonic crystals[12] for optical image processing. For instance, it has been shown that multiple metasurfaces can be used to perform a range of mathematical operations[13] and there have been several theoretical proposals for image differentiation using single layer nanophotonic materials[14–16]. Furthermore, image differentiation has been experimentally demonstrated using photonic crystals[17], the spin hall effect[18], surface plasmon-based devices[19] and Pancharatnam–Berry (PB) phase[20]. However, these past approaches have been restricted to one dimension and some require additional refractive elements (i.e., prisms or lenses) for either plasmon coupling or performing a Fourier transform, which negates the advantage of thin and flat elements.

In this work, we experimentally demonstrate a two-dimensional spatial differentiator that operates in transmission. As shown in Fig. 1a, the differentiator consists of a silicon (Si) nanorod photonic crystal that can transform an image, $I_{in}$, into its second order derivative, $I_{out} = \nabla^2 I_{in}$, allowing for direct imaging of the edges in the image. The use of a 2D photonic crystal allows for differentiation and edge detection in all directions with an NA up to 0.315 and with an experimental resolution smaller than 4 µm. The nanophotonic differentiator is directly integrated into an optical microscope and onto a camera sensor demonstrating the ease at which it can be vertically integrated into existing imaging systems. Furthermore, we demonstrate integration with a metalens for realizing a compact and monolithic image processing system. In all cases, the use of the nanophotonic differentiator allows for a significant reduction in size compared to traditional systems, opening new doors for optical analog image processing in applications involving machine vision.

In order to perform spatial differentiation, an optical filter or material should act as a Laplacian operator on the transmitted light with an intensity profile given by, $I_{out} = \nabla^2 I_{in}$, where $\nabla^2$ is given by $\frac{\partial^2}{\partial x^2} + \frac{\partial^2}{\partial y^2}$. In this case the optical transfer function $H(k_x, k_y)$ follows the function[21]:

$$H(k_x, k_y) = \begin{bmatrix} H_{ss}(k_x, k_y) & H_{sp}(k_x, k_y) \\ H_{ps}(k_x, k_y) & H_{pp}(k_x, k_y) \end{bmatrix} \quad (1)$$

$$= \begin{bmatrix} c_{ss}(k_x^2 + k_y^2) & c_{sp}(k_x^2 + k_y^2) \\ c_{ps}(k_x^2 + k_y^2) & c_{pp}(k_x^2 + k_y^2) \end{bmatrix}$$

where $s, p$ denote the polarization of the incident light and $c_{sp}$ and $c_{ps}$ correspond to polarization conversion. To achieve the required transfer function we employ a photonic crystal slab supporting quasi-guided modes. Unlike modes guided below the light line[22], quasi-guided modes are leaky, propagating in the high-index dielectric slab within the light cone[23,24]. When the frequency and in-plane wave vector match with the quasi-guided modes, Fano interference occurs between the direct transmission and quasi-guided mode which can lead to near-unity back reflection or transmission[25–27]. Fano interference has recently been employed for spatial differentiation using 1D photonic crystals[17]. Here, we employ a 2D design and focus on how the system can be integrated into traditional and nontraditional imaging systems for realizing compact optical analog image processing.

In order to realize the required transfer function we employ a 2D photonic crystal composed of cylindrical silicon (Si) nanorods embedded in polymethyl methacrylate (PMMA) on a silicon dioxide ($SiO_2$) substrate (see Figs. 1a,b). The nanorods have a height of 440 nm, diameter of 280 nm, and period of 600 nm. Fig. 1c shows the simulated color-coded transmittance $|t(f,\theta)|$ as a function of frequency and incident angle ($\theta_{air}$) along the $\Gamma - X$ direction for $s$ and $p$ polarization. At normal incidence, two broad transmission dips are observed at 244 and 268 THz, indicating the presence of two quasi-guided modes with low quality factor. At oblique incidence, the transmission follows a different trend for $s$ and $p$ polarization. Under $s$ polarization, it is observed that the resonant frequencies remain unchanged as a function of incident angle, leading to the matrix components $c_{ss}, c_{ss}$ and $c_{ss}$ being zero in equation (1). However, with $p$ polarization there is a rapid change in transmittance as the incident angle becomes larger. The modulation transfer function $|H(k_x)|$ and phase along the $\Gamma - X$ direction for $p$ polarization are shown in Fig. 1d for a working frequency of 268 THz ($\lambda_0 = 1120$ nm). Importantly, the modulation transfer function has the required quadratic dependence given by $(k_x) = c_{pp} k_x^2$, evidenced by a fit to the simulated data. The quadratic curve is a near perfect fit up to an NA ($nk_x$) of 0.315, which equates to an edge resolution on the scale of 2.17 µm (1.94$\lambda$).

To better understand the polarization dependence, we examined the mode profiles as a function of incident angle. The structure was modeled using a finite difference time domain (FDTD) solver (MEEP[28]) by placing an electric dipole (red dot) within the slab and the in-plane wave vector was swept long the $\Gamma - X$ direction, as shown in Fig. 1e (see details of the simulation methods in Supplementary section 1). Even and odd modes, with respect to the vertical mirror plane (y = 0), are excited by the $E_x$ and $E_y$ components, respectively. From free space, the s (p) polarization can only couple to odd (even) modes due to the field symmetry. While the two modes are degenerate at the $\Gamma$ point ($k_x = 0$), away from normal incidence it can been observed from the $E_z$ profiles in Fig. 1f that the even mode becomes strongly leaky while the odd mode remains confined within the slab (see extended simulation at various $k_x$ in Supplementary Fig. S1). The invariant mode confinement indicates a phase matching condition for the odd mode at various angles, resulting in complete reflection for s polarization. For the even modes, the increased energy leakage leads to an angularly-dependent transmission for p polarization.

In order to experimentally validate the design, we fabricated a $200 \times 200$ µm$^2$ differentiator using electron beam lithography (EBL) in conjunction with reactive ion etching (RIE). A scanning electron microscope (SEM) image of the fabricated device is shown in Fig. 2a. The simulated transmission spectra are presented in Fig.2b and Fig. 2c for s and p polarization, respectively and a schematic of the measurement setup is shown in Fig. 2d. To acquire angle dependent transmission measurements the sample was mounted on a rotation stage and imaged through a magnification system composed of an objective paired with a tube lens. Fig. 2e and f show the measured transmission spectra at various incident angles under s and p polarization, respectively. The incident light is along the $\Gamma - X$ direction with the angles ranging from 0° to 24°. The trend and shape of the measured spectra are in good agreement with the simulation. To map the entire transfer function in *k*-space, we carried out Fourier plane imaging of the device. The nanophotonic differentiator was illuminated by unpolarized light at a wavelength of 1120 nm and a 50X objective (NA=0.42) was used as a condenser with the Fourier images acquired in the back focal plane of a 20X objective (NA=0.4). Fig. 2g and h show the measured back focal plane imaging, along with the transfer function $|H(k)|$ along the $\varphi = 0\,°$ ($\Gamma - X$) and $\varphi = 45\,°$ azimuthal plane ($\Gamma - M$). Along the $\Gamma - X$ direction, the transfer function matches with the fitted parabolic curve over an NA of 0.305. While the Fourier imaging indicates a non-isotropic transfer function, the Laplacian transform at $\varphi = 45°$ can still be fitted with a quadratic function up to an NA of 0.28.

To experimentally quantify the resolution, we used the nanophotonic differentiator to detect the edges of 1951 USAF resolution test chart. A schematic of the imaging setup is shown in Fig. 3a. The test chart was illuminated using unpolarized and collimated light with a wavelength of 1120 nm. The differentiator was placed directly in the front of the test chart which was then imaged through a magnification system comprising an objective paired with a tube lens and a near-infrared camera. The imaging results without the differentiator for element sizes ranging from 30 μm to 4 μm are shown in Fig. 3b. Fig. 3c shows the images of the test chart after being passed through the differentiator. The edges of the micron-scale elements are clearly revealed along both horizontal and vertical directions, which indicates 2D spatial differentiation with a resolution higher than 4 μm. It's also important to note that the differentiator can operate over a relatively broadband due to the low-quality factor resonance. While the differentiator is not an ideal

Laplacian away from the designed wavelength the images in Fig. 3d indicate that it can still operate for edge discrimination across a bandwidth from 1100 and 1180 nm.

One of the primary benefit of flat optics is the ability to vertically integrate them with traditional optical systems. To demonstrate the potential of this approach, we built an edge detection microscope by integrating the image differentiator with a commercial optical microscope (Axio Vert.A1). In this case, the nanophotonic differentiator was redesigned for a wavelength of 740 nm using pillars with a diameter of 180 nm, a period of 385 nm, and a height of 280 nm. Fig. 4a shows a schematic of the microscope setup. The differentiator has a size of $3.5 \times 3.5$ mm$^2$ and is placed below the sample stage directly on top of the microscope objective (10X). An unpolarized monochromatic laser ($\lambda_0 = 740$ nm) was used as the light source incident from the top and imaged on a CCD (uEye). Three types of biological cells were used as the imaging specimen. Fig. 4b-d show the imaging and edge detection results of onion epidermis (b), pumpkin stem (c), pig motor nerve (d). The unfiltered images were obtained at a wavelength of $\lambda = 900$ nm, which is away from the quasi-guided resonance. It can be seen that the shapes and boundaries of cells are less discernable due to the transparent nature of the specimen. By switching to the working wavelength of $\lambda_0 = 740$ nm, we observe clear and high-contrast cell boundaries shown on the right. Such image enhancement mimics the function of phase-contrast microscopy but with significantly reduced system complexity.

Another way in which these filters can be used in traditional optical systems is integration onto a camera sensor. In this case, and in most practical machine vision applications, fabrication at much larger scales is necessary. One potential avenue for scale-up is to employ self-assembly based nanosphere lithography which takes advantage of the inherent periodicity and cylindrical unit cell geometry employed here. We have recently employed this method for realizing large area reflectors and Fig. 5a shows a schematic flowchart of this fabrication process[29]. To investigate the feasibility of this technique for realizing the differentiator we redesigned the device for an operational wavelength of 1450 nm, which corresponds to a rod diameter of 340 nm, height of 480 nm, and a hexagonal lattice with a period of 740 nm. This redesign was necessary to match the periodicity with the size of commercially available nanospheres. The fabrication technique, outlined in detail in the Methods, involves using an array of self-assembled nanospheres as an etch mask for the photonic crystal. Fig. 5b shows the optical image of a fabricated ~1 cm x 1 cm size image differentiator. The color variation corresponds to different grain orientations which does not affect the transmission at different incident angles (see details of transmission map in Supplementary Fig. S3). The SEM images in Fig. 5c indicate high quality Si resonators and a well-defined hexagonal lattice over a large area.

In order to mimic a configuration that may be found in a machine vision application, the large-scale spatial differentiator was placed directly in front of a NIR camera detector, after the imaging lens, as shown in Fig. 5d. For imaging we used transparent centimeter-size plastic flower molds (Figs. 5e,g) as the targets due to their curved surfaces which scatter light at large angles. Figs. 5f,h show the imaging results with and without the filter for two separate objects. Compared to the unfiltered images, the edges of the flowers are clearly revealed when applying the differentiator. While we have not placed the differentiator directly on the sensor in this case, there is nothing that would prevent this in creating a monolithic edge-detecting sensor for machine vision applications.

Lastly, while we have showcased vertical integration with convention optics, the imaging system can be further compacted by employing a metalens as the focusing element for realizing an ultrathin and monolithic image processing system. To create the device we employed multilayer metasurface transfer techniques that we have previously employed for creating doublet lenses[30] and other multilayer metaoptics[31]. Briefly, the metalens and differentiator were designed for operation at λ = 1200 nm and fabricated on separate wafers with the sizes of $200 \times 200$ μm$^2$ followed by embedding in polydimethylsiloxane (PDMS). Fig. 6a and b show the optical images of the differentiator and metalens, respectively, before transfer. The differentiator layer was then released from the handle wafer and transferred on top of the metalens, forming a compound monolithic element shown in Fig. 6c (see details of the fabrication steps in Supplementary section 3 and Fig. S4). The compound metaoptic was then used to image a micron-scale target. Since the target and metalens are small, the images formed by the metaoptic were magnified and reimaged by a 50X objective paired with a tube lens (f=200 mm). Fig. 6d shows the imaging results for wavelengths of 1100 and 1200 nm. At the off-resonant wavelength (λ = 1100 nm) the images are formed without angular filtering while at resonance (λ = 1200 nm) the edges become clearly visible for each of the numbers in the image. In this case there is more noise in the filtered image compared to the experiments employing the filter on a traditional lens or camera sensor. We believe that the additional noise can be attributed to reflections between the object and differentiator layer which is stronger in this case due to the short focal length of the lens. This issue can be minimized by making a larger lens with a longer focal length.

In conclusion, we have experimentally demonstrated two-dimensional image differentiators with high resolution, thin form factor, and a simple geometry which allows rapid and cost-effective large-scale manufacturing. Furthermore, we have demonstrated how a complete image processing system can be accomplished using monolithic compound flat optics. These types of optical analog image processors could open new doors for applications in areas such as biological imaging and machine vision. Metaoptics with more complex *k*-space response could also be realized by employing multilayer architectures[30] and inverse design mechanisms[32–34] for applications such as more complex optical analog computing and augmented reality displays.

## Materials and methods

### Simulations

The transmission spectra were calculated using the Frequency Domain (FD) solver of CST Microwave Studio. The refractive index of $SiO_2$ and PMMA were set to be 1.45 and 1.48, respectively, and the index of Si (3.67 at 1120 nm) was obtained using ellipsometry. The Si nanorods were modeled as a periodic unit cell on a $SiO_2$ substrate embedded in a PMMA cladding layer. The details of the quasi-guided resonance simulations are provided in Supplementary section 1.

### Fabrication

The photonic crystal slab was defined in a 440 nm thick a-Si layer that was grown on a $SiO_2$ substrate using low pressure chemical vapor deposition (LPCVD). A 200 nm PMMA A4 layer was spin coated at 4500 rpm followed by the deposition of 10 nm thick chromium as the conduction layer using thermal evaporation. The patterns were then defined using electron beam lithography followed by depositing 35 nm thick alumina oxide as a dry etch mask using e-beam evaporation. The Si nanorods were then etched using reactive ion etching using a mixture of $C_4F_8$ and $SF_6$, Finally, five layers of PMMA A4 were sequentially spin coated to encapsulate the Si nanorods. The same procedure was also used for fabricating the visible image differentiator.

In the case of nanosphere lithography a monolayer of 740 nm diameter nanospheres was first formed at the water-air interface of a bath through controlled injection by a syringe pump. A wafer with a 480 nm thick layer of Si on SiO$_2$ was titled at 10° and placed at the bottom of the Teflon bath. The densely packed hexagonal nanospheres were transferred to the wafer by slowly draining the bath. The nanospheres were then downscaled using an O$_2$ plasma and used as an etch mask during reactive ion etching to define the Si rods. The fabrication details of the compound metaoptics and additional details regarding nanosphere lithography are provided in Supplementary sections 2 and 3.

## Acknowledgements

The authors acknowledge support received from the Office of Naval Research under award N00014-18-1-2563 and DARPA under the NLM program, award HR001118C0015. Part of the fabrication process was conducted at the Center for Nanophase Materials Sciences, which is a DOE Office of Science User Facility.

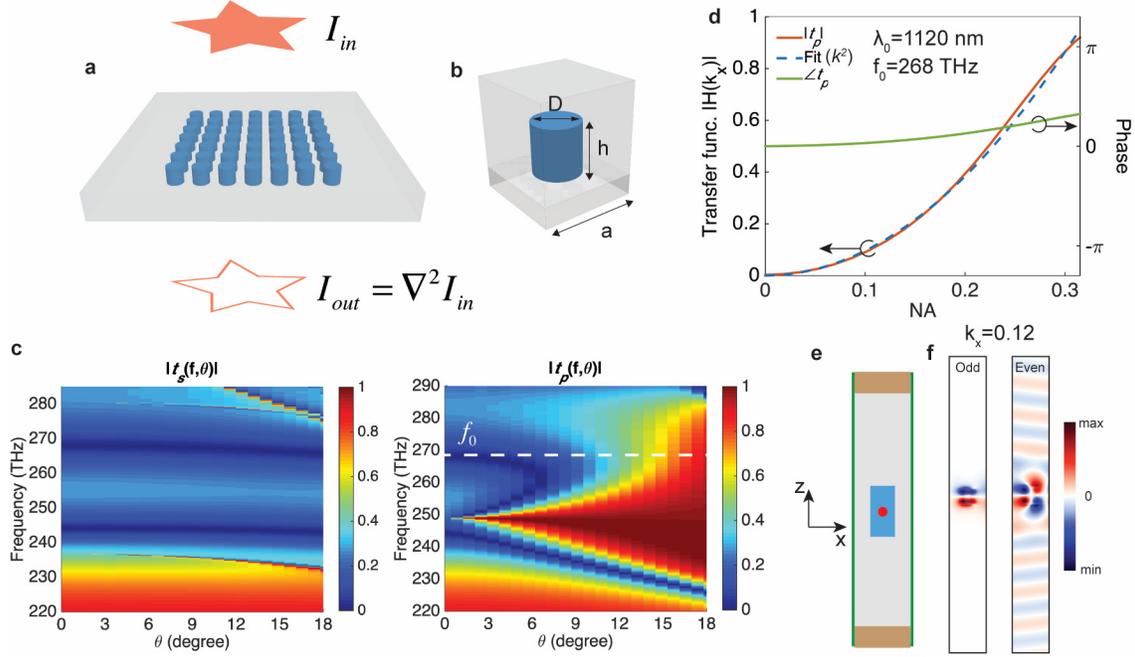

**Fig. 1 | Two-dimensional image differentiation using nanophotonic materials.** (a) Schematic of a dielectric nanophotonic slab acting as a Laplacian operator that transforms an image, $I_{in}$, into its second order derivative, $I_{out} = \nabla^2 I_{in}$. (b) Unit cell of the nanoarray composed of Si nanorods. The array is embedded in a layer of polymethyl methacrylate (PMMA) on a SiO$_2$ substrate. (c) Simulated color-coded transmittance $|t(f, \theta)|$ as a function of frequency and incident angle ($\theta$) along the $\Gamma - X$ direction ($\varphi = 0°$) for $s$ and $p$ polarization. (d) Optical transfer function $H(k_x)$ along the $\Gamma - X$ direction for $p$ polarization at $\lambda_0 = 1120$ nm, and the quadratic fitting in the form of $c_{pp} k_x^2$. (e) Schematic of the simulation model for the quasi-guided modes. An electric dipole (red dot) was placed within the slab as an excitation source. (f) $E_z$ field distributions showing different mode excitations at $k_x = 0.12$ ($2\pi/a$) for odd and even modes.

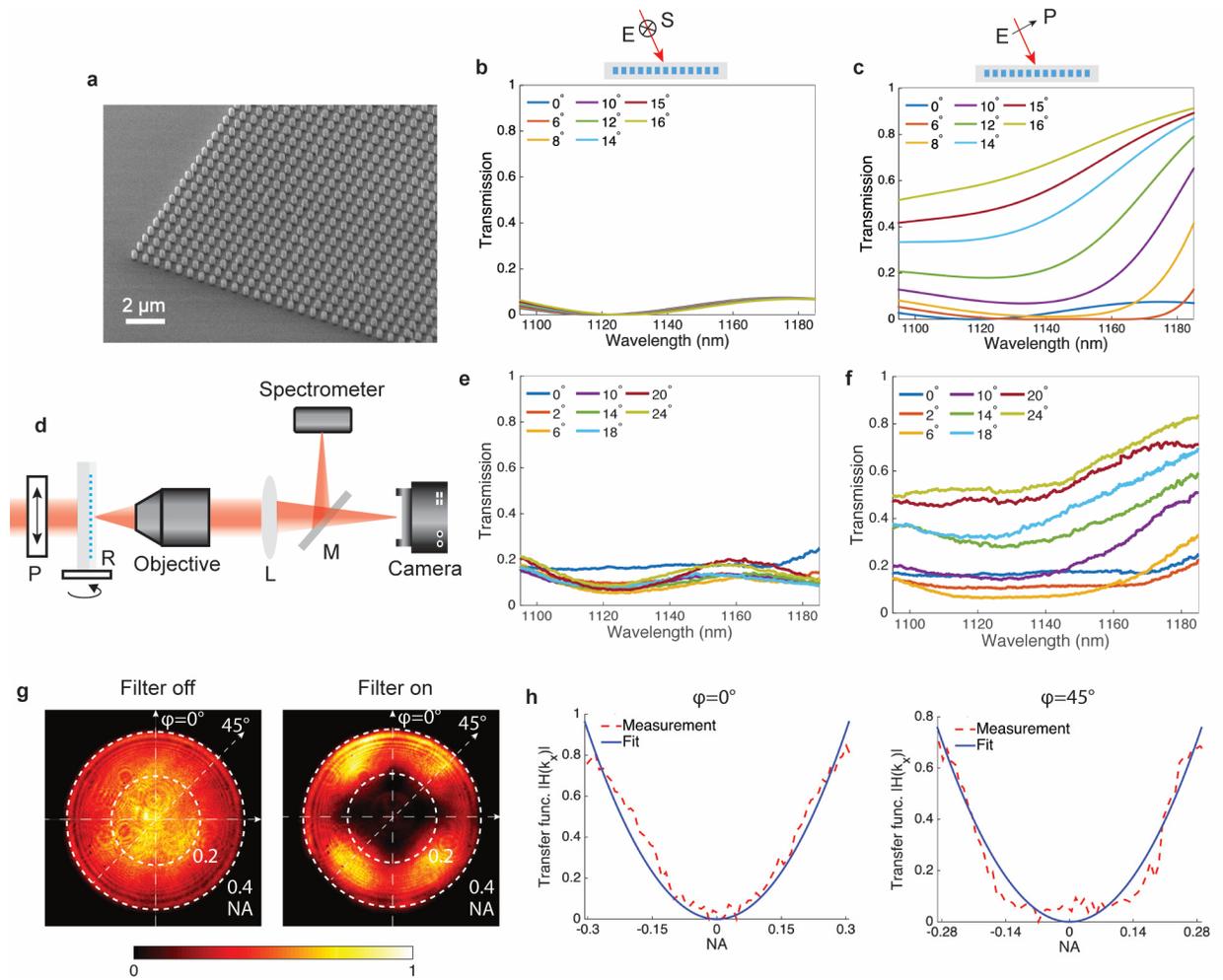

**Fig. 2 | Fabrication and characterization of the nanophotonic spatial differentiator.** (a) SEM image of the fabricated Si photonic crystal. (b) Experimental setup for measuring the transmission spectra at various angles. P, polarizer; R, rotation stage; L, tube lens (f=200 mm); M, flip mirror. (c-d) Simulated transmission spectra along the $\Gamma - X$ direction for $p$ (c) and $s$ (d) polarization. (e-f) Measured transmission spectra for $p$ (c) and $s$ (d) polarized incident light. (g) Measured back focal plane images without and with the nanophotonic differentiator. (h) Extracted 1D modulated transfer function along $\varphi = 0°$ and 45°.

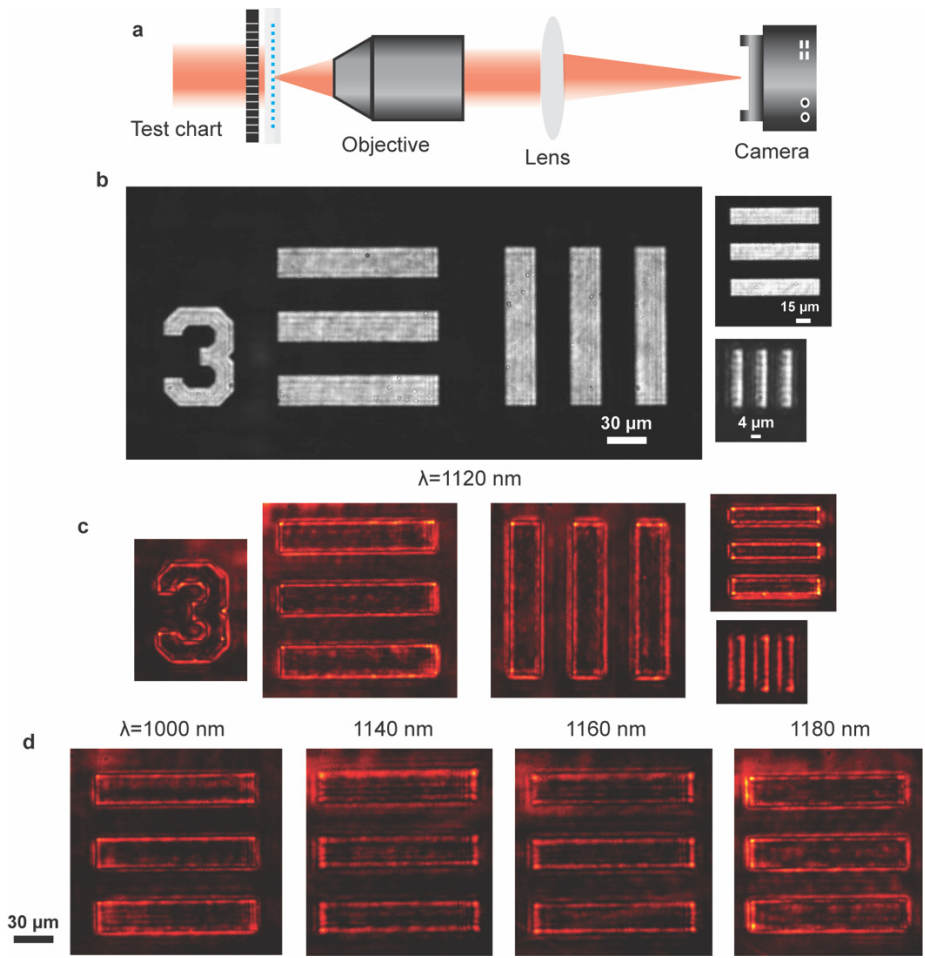

**Fig. 3|Differentiator resolution characterization.** (a) A schematic of the imaging setup. The nanophotonic differentiator is placed in front of a standard 1951 USAF test chart and the targets are magnified through an objective paired with a tube lens. (b-c) Imaging results of the target without (b) and with (c) the filter. (d) Edge detection results at different wavelengths ranging from $\lambda_0 = 1100$ nm to 1180 nm.

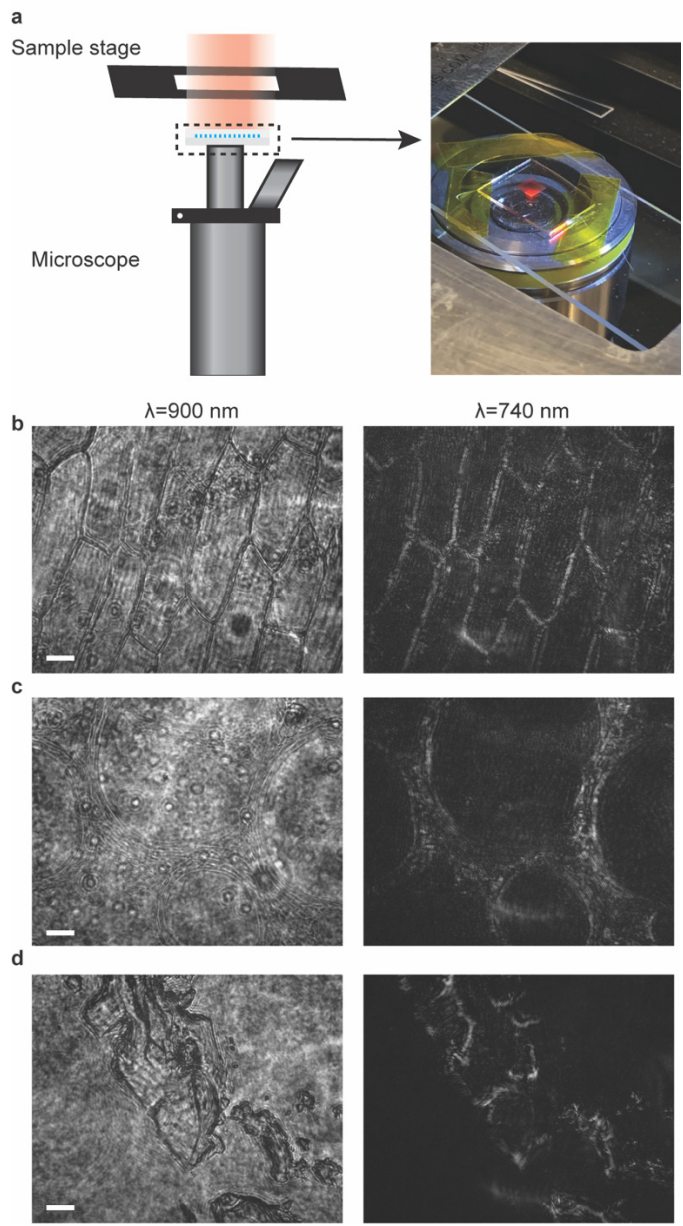

**Fig. 4 | Edge detection microscope at visible frequencies.** (a) A schematic of the edge detection microscope. The spatial differentiator is redesigned at the wavelength of $\lambda_0 = 740$ nm and fabricated at a size of $3.5 \times 3.5$ mm$^2$, which is directly integrated with a commercial inverted optical microscope (Axio Vert.A1). (b-d) Imaging and edge detection results of three types of biological cell samples. (b), onion epidermis; (c), pumpkin stem; (d), pig motor nerve. Images on the left are obtained at the wavelength of $\lambda = 900$ nm which are off the quasi-guided resonant frequency, and the images on the right correspond to the results at the working wavelength of $\lambda_0 = 740$ nm. Scale bar: 50 μm.

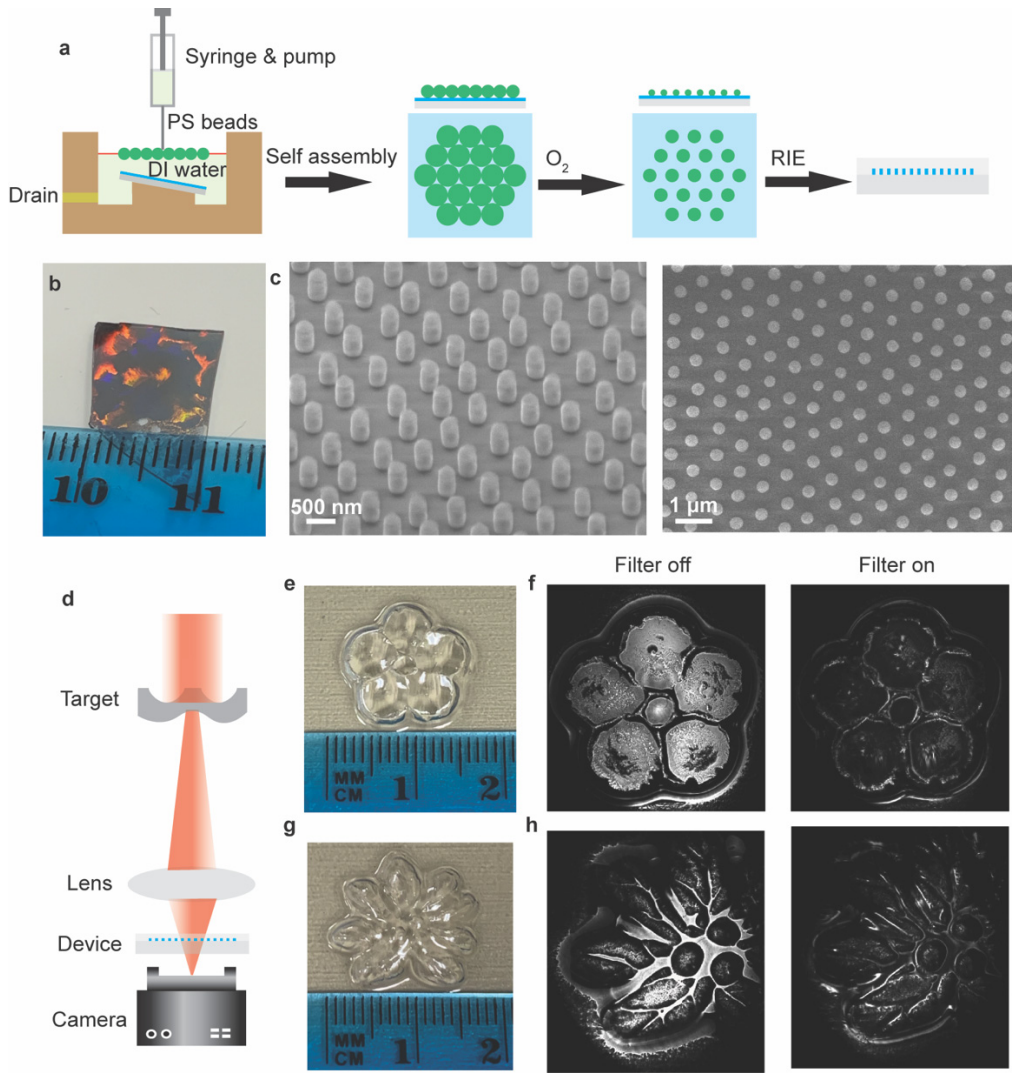

**Fig. 5 | Large-scale nanophotonics using nanosphere lithography.** (a) Flowchart of the fabrication process. A monolayer of nanospheres (diameter of 740 nm) were formed at the water-air interface of a bath and then transferred to a tilted substrate with a Si film (thickness of 480 nm), resulting in a densely-packed nanospheres arranged in a hexagonal lattice. The nanospheres were then downsized and used as a dry etch mask for defining the Si nanostructures. (b) Optical image of a centimeter-scale spatial differentiator. (c) SEM images of the Si rods. The device is designed at a wavelength $\lambda_0 = 1450$ nm. (d) Schematic of the imaging setup. The large-scale device is placed in front of a NIR camera sensor. (e) Optical image of a plastic flower mold which was used a 3D macroscopic imaging target. The size of the object is on the scale of centimeters. (f) Imaging and edge detection results. Images on the left and right correspond to the systems without and with the angular differentiator, respectively. (g-h) The same imaging results on a second target.

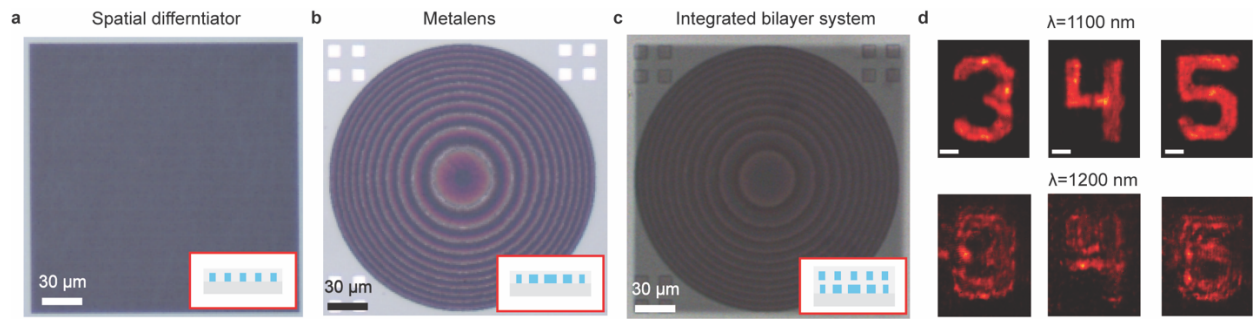

**Fig. 6|Compound metaopitc.** (a-c) Optical images of the nanophotonic differentiator (a), metalens (b) and monolithic compound system (c). The insets correspond to schematics of the device cross-sections. (d) Imaging results for the off-resonant wavelength of 1100 nm (top) and the working wavelength of 1200 nm (bottom). Scale bar: 20 μm.